\documentclass[pra,twocolumn,showpacs,preprintnumbers,amsmath,amssymb,superscriptaddress]{revtex4}
\usepackage{graphicx}
\usepackage{dcolumn}
\usepackage{bm}
\usepackage{stmaryrd}
\usepackage{latexsym}
\usepackage{amssymb}
\usepackage{amsfonts}
\usepackage{amsmath}

\begin{document}

\title{Phase separation in optical lattices in a spin-dependent external potential}
\author{A-Hai Chen}
\affiliation{Department of Physics, Zhejiang Normal University, Jinhua 340012, China}
\author{Gao Xianlong}
\email{gaoxl@zjnu.edu.cn}
\affiliation{Department of Physics, Zhejiang Normal University, Jinhua 340012, China}
\affiliation{Kavli Institute for Theoretical Physics China, CAS, Beijing 100190, China}
\date{\today}
\begin{abstract}

We investigate the phase separation in one-dimensional Fermi gases on optical lattices. The density distributions and the magnetization are calculated by means of density-matrix renormalization method. The phase separation between spin-up and spin-down atoms is induced by the interplay of the spin-dependent harmonic confinement and the strong repulsive interaction between intercomponent fermions. We find the existence of a critical repulsive interaction strength above which the phase separation evolves. By increasing the trap imbalance, the composite phase of Mott-insulating core is changed into the one of ferromagnetic insulating core, which is incompressible and originates from the Pauli exclusion principle.
\hspace*{7.0cm}
\end{abstract}


\maketitle
\section{Introduction}
Ultracold atoms in optical lattices provide a new test bed for interacting quantum many-body systems~\cite{review paper}. Fermionic atoms in optical lattices can be used to realize the clean Fermi-Hubbard model, which is free of lattice defects, impurities, and phonons, in contrast to those in solid-state systems. Over the past few years, many interesting phenomena were observed in optical lattices, for example, the Fermi surface of the atoms in the lattice, the transform from a normal state into a band insulator~\cite{Kohl}, and fermionic superfluidity of attractively interacting fermions~\cite{Chin}. Two other major breakthroughs achieved recently in fermionic superfluid are: the BEC-BCS crossover~\cite{Regal} and imbalanced superfluidity~\cite{Zwierlein}.

The spatial inhomogeneity due to the confinement essential for ultracold atomic experiments is always present, which leads to a spatially varying local density distribution and normally invalidates a reliable analytical method usually used in the homogeneous system. Many numerical schemes such as the density-matrix renormalization group (DMRG)~\cite{Machida1,gaoprl98,Molina}, quantum Monte Carlo~\cite{Rigol,Rigol2004,Astrakhardik,Casula}, exact diagnolization~\cite{exact diagnolization,Machida1,Nikkarila}, and density-functional theory based on the exact Bethe-ansatz solution~\cite{gaoprl98} are used in studying the many interesting quantum effects in spin-balanced or imbalanced systems. Among them, intriguing properties such as phase separation in a trap and the transition from superfluidity to a normal state have attracted a great deal of attention both experimentally and theoretically.

In the experiments, a phase separation was observed between the normal component and the superfluidity of interacting fermionic atom gases with imbalanced spin populations~\cite{Zwierlein}. In theory, the mean-field approach provides a qualitative explanation of the phase separation of imbalanced fermionic atom gases in a trap~\cite{mean-field theory}. The imbalance of the two species with $N_\uparrow\ne N_\downarrow$ can be produced by different trapping frequencies~\cite{Recati06}, namely, spin-dependent trapping potentials. Phase separation can occur in trapped spinor boson gases with a weak anisotropic spin-spin interaction~\cite{Hao} and in multicomponent Fermi gases with different values of the scattering lengths and particle number~\cite{Roth}. In two-dimensional optical lattices, the phase separation due to the imbalanced mixture, antiferromagnetic order~\cite{Snoek}, and pairing symmetries~\cite{An} is investigated by the mean-field theory. For a one-dimensional (1D) system of two-component Fermi gases in a continuous space, it is found that there exists a critical interaction strength beyond which one atomic component expels another from the center of the trap~\cite{Karpiuk}. For a 1D attractive Hubbard model, a phase separation between the condensate and unpaired majority atoms occurs for a certain range of the interaction and population imbalance. At $T=0$ beyond a critical spin polarization, the phase separation always exists no matter how strong the interaction is~\cite{Tezuka}. For a 1D repulsive Hubbard model, the phase separation due to the different trap frequencies is discussed within the local magnetization by the spin-dependent density-functional theory~\cite{Abedinpour}.

In the present work, we are interested in the phase separation between different fermion species induced by the spin-dependent external potentials. The interplay between the external spin-dependent potentials and the repulsive interaction of intercomponent fermions will be explored.

\section{The model}
We consider a two-component Fermi gas in a tube with $N_f$ atoms and $N_s$ lattice sites with the unit lattice constant, which can be described by a one-band inhomogeneous Fermi-Hubbard model~\cite{jaksch_98}
\begin{eqnarray}\label{eq:hubbard}
\hat {H}_s&=&-t\sum_{i,\sigma}^{}({\hat
c}^{\dagger}_{i\sigma}{\hat c}_{i+1\sigma}+{H}.{c}.)+
U\sum_{i=1}^{N_s}\,{\hat n}_{i\uparrow}{\hat n}_{i\downarrow}\nonumber\\
&&+\sum_{i,\sigma}^{}V_\sigma\left[i-(N_{s}-1)/2\right]^2{\hat n}_{i}\,,
\end{eqnarray}
where the spin degrees of freedom $\sigma=\uparrow,\downarrow$ are pseudospin-$1/2$ labels for
two internal hyperfine states and ${\hat c}_{i\sigma}$ (${\hat c}^{\dagger}_{i\sigma}$) are fermionic operators annihilating (creating) particles with spin $\sigma$ in a Wannier state at site $i$. ${\hat n}_i= \sum_{\sigma} {\hat n}_{i\sigma}=\sum_{\sigma} {\hat c}^{\dagger}_{i\sigma}{\hat c}_{i\sigma}$ is the total site occupation operator, $t$ is
the tunneling between the nearest neighbors, $U$ is the strength of the on-site interaction, and $V_\sigma$ describes the strength of the spin-dependent harmonic trapping potentials $V_{\rm har, \sigma}=V_\sigma\left[i-(N_{s}-1)/2\right]^2$.

The inhomogeneous Fermi-Hubbard model can be realized by a strong confinement in transverse directions~\cite{moritz_2005} with an additional periodic potential applied along the tube. Concerning the experimental realization of the spin-dependent external potentials, one can use magnetically trapped Fermi mixtures of a particular atom in the two different hyperfine states~\cite{equalmass,Iskin}, or two different trapped atoms of unequal masses~\cite{Blume}, where the different magnetic moments make $V_\uparrow\ne V_\downarrow$. In the experiment of two $^{40}$K fermion species, the ratio of frequencies $V_\uparrow/V_\downarrow=\sqrt{9/7}$ is discussed~\cite{Jin}.
In optical lattices, a spin-dependent optical trap can be realized by asymmetrically detuning the laser frequencies with respect
to the two hyperfine states~\cite{Iskin}. Experimentally, the atomic density we calculated is the most convenient and clear observable detectable by electron beams, high-resolution cameras, or noise interference. Recently, a composite phase of an incompressible Mott-insulator phase in the core was identified~\cite{Jordens}, where the core is composed of strongly repulsive fermionic atoms in two hyperfine states. It is shown how the system evolves by increasing confinement from a compressible dilute metal into a band-insulating state, which also provides a way to polarize a spin-balanced system where $N_\uparrow=N_\downarrow=N_f/2$ ~\cite{Jordens}.

The homogeneous 1D Fermi-Hubbard model belongs to the universality class of Luttinger liquids. At zero temperature, the properties of this model in the thermodynamic limit ($N_{\sigma}, N_s\rightarrow \infty$, but with finite $N_{\sigma}/N_s$) are determined by the fillings $n_\sigma=N_\sigma/N_s$ and by the dimensionless coupling constant $u=U/t$. According to Lieb and Wu~\cite{LiebWu}, the ground state (GS) properties for different fillings in the thermodynamic limit are described by the coupled integral equations (for details see Refs. [32,33]).

For the inhomogeneous system described by Eq. (\ref{eq:hubbard}), the coexisted phases induced by the external spin-independent trapping potentials ($V_\uparrow=V_\downarrow$) were well identified by many authors~\cite{Rigol,Rigol2004,Heiselberg,Campo,Liu}. We focus in this work on the spin-dependent potentials ($V_\uparrow\ne V_\downarrow$) by applying the DMRG techniques, performed by using the ALPS libraries~\cite{Albuquerque}. During our DMRG calculations, the states kept are $500$ to $1000$ so that we can restrict the cut error to be less than $10^{-11}$.

\section{Numerical results}
In this section we present our numerical results. In the following discussion we keep the total number of particles constant ($N_f=40$) and vary the number of spin-up and spin-down atoms in the system. We characterize the confinement imbalance by defining the ratio between the spin-up and spin-down dependent external potentials as
\begin{eqnarray}\label{eq:ratio}
\gamma=\frac{V_\uparrow}{V_\downarrow}.
\end{eqnarray}

\begin{figure}
\begin{center}
\includegraphics*[width=1.00\linewidth]{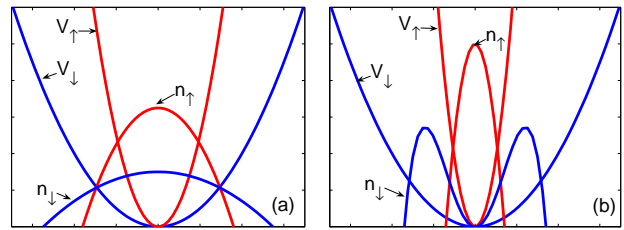}
\caption{(Color online) Schematic illustrations for spin-dependent harmonic potentials $V_{\rm har, \sigma}$ (in units of $t$) and the density distributions $n_\sigma$ (in units of the lattice constant) of both spin-up and spin-down atoms in the presence of interactions. The left panel (a) is for the system of small trap imbalance, where the spin-up and spin-down atom mixture in the center forms a PM region and the right panel (b) for the system of large trap imbalance, where the PS region is formed.
\label{fig:one}}
\end{center}
\end{figure}

In Fig. \ref{fig:one} we show schematic plots for the spin-dependent harmonic potentials and the density distributions of the two fermion species with small or large confinement imbalances. The effects of $\gamma$ are manifested in that two atoms coexist for small $\gamma$ where the spin-up and spin-down atom mixture in the center of the trap forms phase mixing (PM) region and separate with only spin-up atoms left for large $\gamma$ where the phase separated (PS) region is formed. The PS region is determined with the local occupation in the trap center (i.e., $i$=0) satisfying $n_0\le 10^{-3}$. We distinguish in the following the different phases by showing the atomic density profiles and the local magnetization for different repulsive interactions and confining strengths in the system of spin-unpolarized or spin-polarized atoms.
\begin{figure}
\begin{center}
\includegraphics*[width=1.00\linewidth]{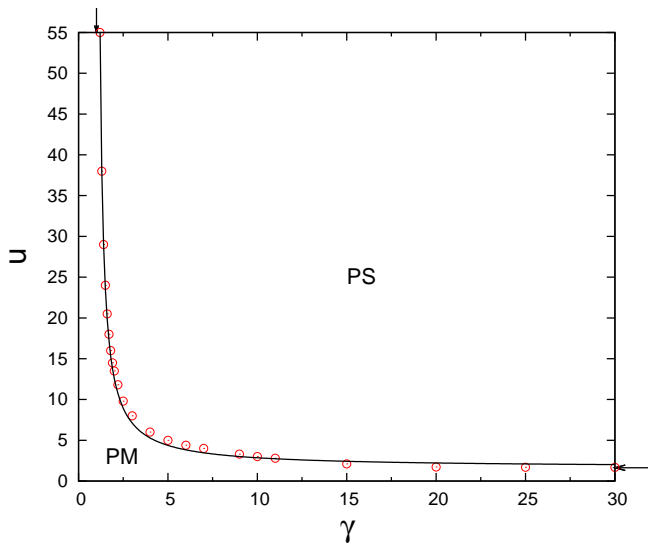}
\caption{(Color online) Phase diagram as a function of $u$ and the confinement ratio $\gamma$ determined by the DMRG technique. The system consists of $N_\uparrow=20$ and $N_\downarrow=20$ fermions. The spin-down trap strength is $V_\downarrow=1.0\times 10^{-3}$. The arrow in the top indicates the position where $\gamma=1$ and the arrow in the right where $u_c=1.64$. The solid line is a power-law fit $u=u_c+\alpha/(\gamma-1)$ to the data with $\alpha=10.932$. The two phases PM and PS are manifested in Fig. \ref{fig:one} and explained in the text.
\label{fig:two}}
\end{center}
\end{figure}

\begin{figure}
\begin{center}
\includegraphics*[width=1.00\linewidth]{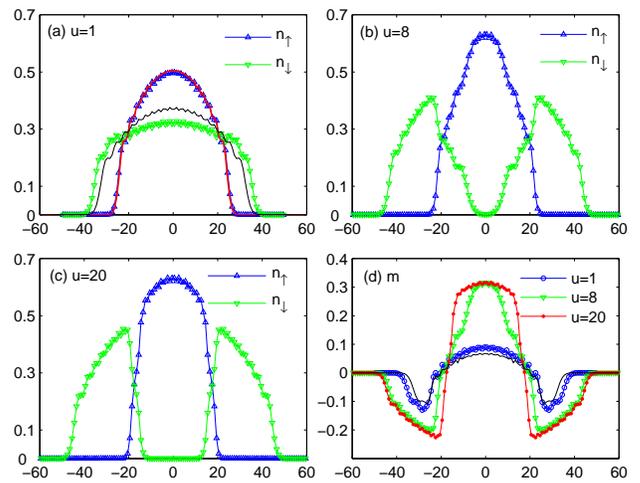}
\caption{(Color online) Ground state density profiles for $n_{i\uparrow}$, $n_{i\downarrow}$ and local magnetization $m_i=(n_{i\uparrow}-n_{i\downarrow})/2$ as a function of $i$ in the spin-dependent external potentials of $\gamma=3$. The spin-down trap strength is $10^{-3}$. The system consists of $N_\uparrow=20$ and $N_\downarrow=20$ fermions. Three different interaction strengths are shown: (a) $u=1$, (b) $u=8$, and (c) $u=20$. The local magnetization $m_i$ is shown in (d). The solid line connecting the symbols serves as a guide for the eyes.
For comparison, the GS densities of the noninteracting case ($u=0$) for spin-up (bold solid line) and spin-down (thin solid line) atoms are included in (a), and the corresponding local magnetization (thin solid line) is also shown in (d).
We find that the repulsive interaction can induce a complete phase separation between the two components in the spin-dependent external potentials.\label{fig:three}}
\end{center}
\end{figure}

First, we study the phase separation between the two-component fermions induced by the interplay between the repulsive interaction and the spin-dependent parabolic potentials in the unpolarized system of an equal number of spin-up and spin-down atoms ($N_\uparrow=N_\downarrow=20$) and $V_\downarrow=1.0\times 10^{-3}$. The lattice size chosen here and in the following is always large enough to make sure that the GS densities smoothly drop down at the edges. In Fig. \ref{fig:two}, the phase diagram is shown as a function of $u$ and the confinement ratio $\gamma$. Two regions are seen: the PM region with both spin-up and spin-down mixtures in the center of the trap and the PS region with only spin-up atoms remaining in the center. A critical interaction strength $u_c=1.64$ is obtained, below which there is no phase separation no matter how large the confinement ratio. For the system considered here, the condition in which the phase separation happens can be simply fitted by a power-law relation $u=u_c+\alpha/(\gamma-1)$, with $\alpha=10.932$.
We further illustrate in Fig. \ref{fig:three} an explicit example by choosing $\gamma=3$ and changing the interaction strength.  We confirm that there exists a critical value of the interaction strength ($u=8$) beyond which the spin-down atoms are depleted from the center of the trap and repelled into the periphery regions between $V_\uparrow$ and $V_\downarrow$. In this case, a phase separation begins to appear (i.e., the Fermi components tend to stay in different spatial regions). Thus, upon approaching the phase separation point and beyond, the local polarization of the atomic gases in the center becomes stronger and stronger. When the complete phase separation is realized, fermions become fully polarized due to the strong repulsive interaction. As a result, spin-up atoms locate in the center and spin-down atoms at the periphery of the trap, which is clearly seen in Fig. \ref{fig:three}(c) for $u=20$. Upon reaching the complete phase separation, further increasing the repulsive interaction only makes the spin-up density a little more confined and spin-down density more spread out. In Fig. \ref{fig:three}(d), we plot the local magnetization of the system, which is defined as $m_i=(n_{i\uparrow}-n_{i\downarrow})/2$. For the strong repulsive interaction where the phase separation begins to evolve $m_i$ changes from negative to positive with a big slope.

\begin{figure}
\begin{center}
\includegraphics*[width=1.00\linewidth]{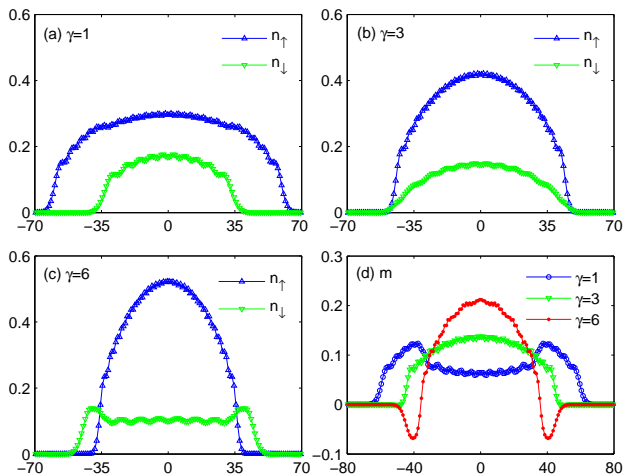}
\caption{(Color online) Ground state density profiles for $n_{i\uparrow}$, $n_{i\downarrow}$ and local magnetization $m_i$ as a function of $i$ for the system of weak repulsive interaction ($u=1$) in the spin-dependent external potentials. The system consists of
$N_\uparrow=30$ and $N_\downarrow=10$ fermions with $V_\downarrow=2.5\times 10^{-4}$. Three different ratios of confining potentials are shown. (a) $\gamma=1$, (b) $\gamma=3$, and (c) $\gamma=6$. The local magnetization $m_i$ is plotted in (d). From (d), we can see that upon reaching the phase separation point and beyond, the local magnetization $m_i$ becomes more negative at the periphery and more positive in the bulk region of the trap signaling that more spin-down fermions are repelled from the center and more spin-up fermions are constrained there.\label{fig:four}}
\end{center}
\end{figure}

\begin{figure}
\begin{center}
\includegraphics[width=1.00\linewidth]{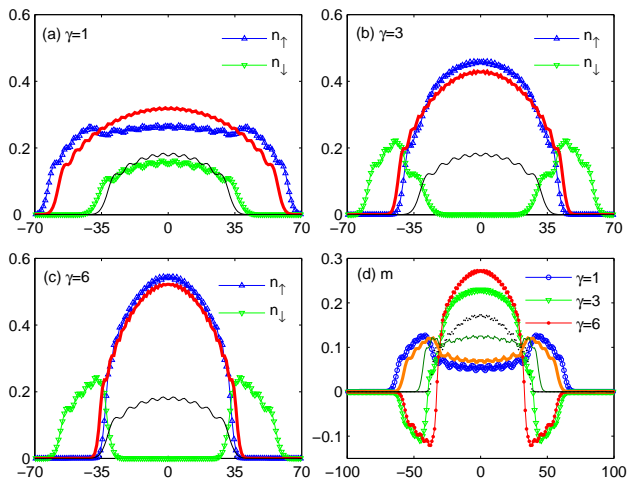}
\caption{(Color online) Same as Fig. \ref{fig:four} but for the system of strong repulsive interaction ($u=4$).
From (b), we notice that a complete phase separation occurs.
For comparison, in (a)-(c), the GS densities of the noninteracting case ($u=0$) for the spin-up (bold solid line) and spin-down (thin solid line) atoms are also plotted. In (d) we include the corresponding local magnetizations for the noninteracting case with $\gamma=1$ (bold solid line), $3$ (thin solid line), and $6$ (dotted line), respectively.
\label{fig:five}}
\end{center}
\end{figure}

Now, let us concentrate on the phase separation induced by spin-dependent parabolic potentials.
In Figs. \ref{fig:four} and \ref{fig:five} we study the polarized systems of an unequal number of spin-up and spin-down particles ($N_\uparrow=30$ and $N_\downarrow=10$) with weak ($u=1$) and strong ($u=4$) repulsive interactions. We illustrate the effects of the confinement ratio on the local density distributions and the local magnetization.
We increase $\gamma$ by keeping the spin-down external potential $V_\downarrow$ as invariant and increasing $V_\uparrow$ (i.e., $\gamma\ge 1$).
In Fig. \ref{fig:four}, the density profiles for the weak repulsive interaction ($u=1$) are shown with different confinement imbalances ($\gamma=1, 3$, and $6$). While increasing the confinement for the spin-up atoms, the interaction between the spin-up and spin-down atoms in the center of the trap repels the spin-down atoms into the edges of the trap. However the repulsive interaction is not strong enough and only a small amount of phase separation appears.
From Fig. \ref{fig:five}(b), we can see that, in the system of strong repulsive interaction ($u=4$) and a large trap imbalance ($\gamma=3$), almost all the spin-down atoms are repelled from the bulk of the trap and a complete phase separation is realized. Due to the depletion of the spin-down fermions, fully polarized gases of spin-up fermions are obtained, as can be seen in Figs. \ref{fig:five}(b) and \ref{fig:five}(c). For comparison, the GS density distributions of the spin-up and spin-down atoms for the noninteracting case ($u=0$) are also included, where no phase separation is observed. We conclude that the intercomponent interaction is essential in achieving a phase separation between the two-component fermions in a spin-dependent trap.

The local magnetization $m_i$, in Figs. \ref{fig:four}(d) and \ref{fig:five}(d), gives another signature of the phase separation. For small $\gamma$, a flat region of $m_i$ is seen in the center of the trap and two bumps are shown at the edges with the excess spin-up atoms. The increase of the trap imbalance and the repulsive interaction strength shows a signature that $m_i$ is more negative at the edges, that is, more and more spin-down atoms are repelled from the center of the trap and accumulate at the periphery region between $V_\uparrow$ and $V_\downarrow$.

\begin{figure}
\begin{center}
\includegraphics*[width=0.70\linewidth]{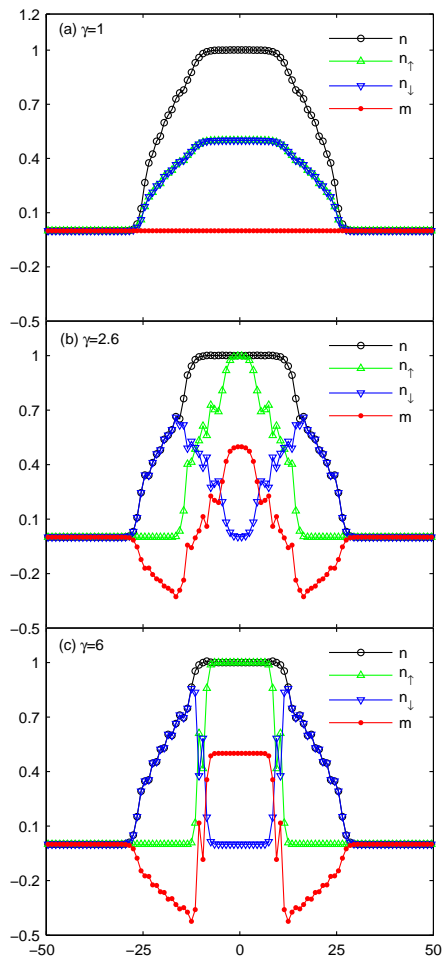}
\caption{(Color online) Density distributions for spin-up and spin-down fermions together with their sum (the total GS density) and difference (the local magnetization) plotted against the site with strong repulsive interaction ($u=6$) in the spin-dependent external potentials. The system consists of $N_\uparrow=20$ and $N_\downarrow=20$ fermions with $V_\downarrow=6.0\times 10^{-3}$. Three different ratios of confining potentials are shown: (a) $\gamma=1$, (b) $\gamma=2.6$, and (c) $\gamma=6$.
\label{fig:six}}
\end{center}
\end{figure}
\begin{figure}
\begin{center}
\includegraphics*[width=0.70\linewidth]{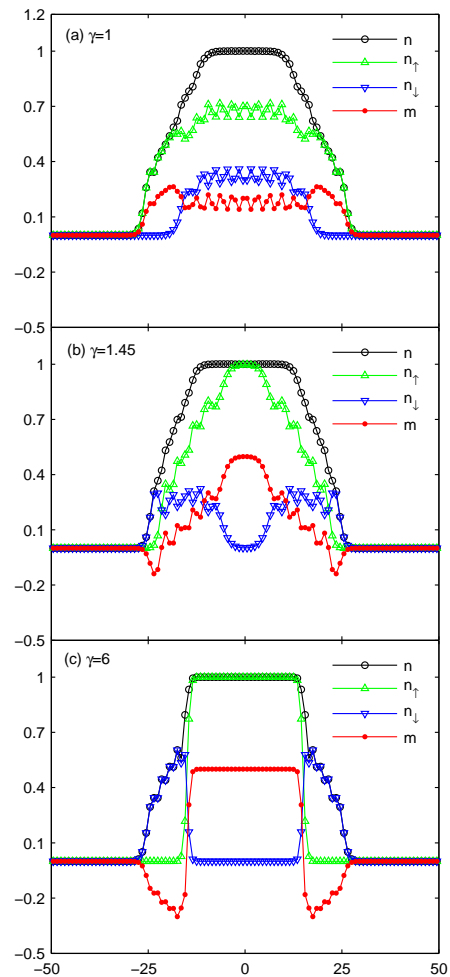}
\caption{(Color online) Same as Fig. \ref{fig:six} but for the spin-polarized system of $N_\uparrow=30$ and $N_\downarrow=10$. Three different ratios of confining potentials are shown: (a) $\gamma=1$, (b) $\gamma=1.45$, and (c) $\gamma=6$.
\label{fig:seven}}
\end{center}
\end{figure}
In the following we study how the spin-dependent potentials influence the composite phase of the Mott-insulating core in the bulk.
In Fig. \ref{fig:six}, we show the GS density distributions of an unpolarized system of $N_\uparrow=N_\downarrow=20$ under the influence of the different trap imbalances with $V_\downarrow=6.0\times 10^{-3}$ and a strong repulsive interaction of $u=6$. For $\gamma=1$, a Mott phase is formed in the bulk region of the trap. With the increase of the confinement for the spin-up atoms, the spin-down atoms are repelled from the center of the trap. The Mott phase induced by the interaction between the locally spin-balanced fermions is changed into the Mott-like phase induced by both the interaction between the locally spin-imbalanced fermions and the spin-dependent potentials. At the critical point of $\gamma=2.6$, the phase separation starts [see Fig. \ref{fig:six}(b)] and the strong confinement for the spin-up atoms forms an insulating core of fully polarized fermions in the center of the trap, over which the local occupancy is a unit. This insulating core is regarded as a ferromagnetic insulating phase since it is incompressible in nature and originates from the Pauli exclusion principle, which differs from the Mott-insulating phase induced by the repulsive interaction between fermions~\cite{MachidaPRB}, such as in Fig. \ref{fig:six}(a). The unit core becomes stable for $\gamma>6$ by further increasing $\gamma$. Upon reaching the phase separation point and beyond, a plateau of constant $m_i=0.5$ is formed in the center of the local magnetization.

In Fig. \ref{fig:seven} we show the case of a polarized system of $N_\uparrow=30$ and $N_\downarrow=10$ with a strong repulsive interaction of $u=6$. For $\gamma=1$, the spin-up fermions form Wigner-lattice-type profiles inside the Mott core~\cite{Soffing}, which occurs at low fillings or equivalently at large $u$ and can be explained by mapping Eq. (\ref{eq:hubbard}) into the antiferromagnetic Heisenberg model~\cite{MachidaPRB}. We notice that, compared to the unpolarized system of that in Fig. \ref{fig:six}, the polarized system becomes more easily reaches the phase separation point ($\gamma=1.45$). That is, the Mott phase in the polarized system is less robust
against the increase of the interaction strength and confinement.

\section{Conclusions}
In this article we perform a theoretical study of a 1D Fermi-Hubbard model in a spin-dependent harmonic trap within the DMRG techniques. The interplay between the repulsive interaction and the spin-dependent harmonic trap is studied for the system of spin-balanced or spin-imbalanced Fermi gases. We find that, for the system in the spin-dependent external potentials, there exists a critical interaction strength beyond which a phase separation can occur with two Fermi components staying in the different spatial regions. For the system with a weak interaction strength, upon increasing the trap imbalance, the spin-up atoms are confined more and more in the center of the trap and a depletion occurs for the spin-down atoms due to the intercomponent repulsive interactions. However, the weak repulsive interaction below a critical value is not capable of achieving a full phase separation. For the system with strong intercomponent repulsive interactions, a complete phase separation is realized at the strong confinement imbalance where spin-down atoms are repelled out of the bulk region with only spin-up atoms remaining.

For the system with both strong confinement and strong repulsive interactions, where a composite phase of the Mott-insulating core is formed in the center, we show that, upon increasing the trap imbalance, the Mott phase induced by the interaction between the locally spin-balanced fermions is changed into the Mott-like phase induced under the interplay between the interaction of the locally spin-imbalanced fermions and the spin-dependent confining potentials. Upon reaching the phase separation point and beyond, the ferromagnetic insulating phase due to the Pauli exclusion principle appears, which is of the unit core. In the distribution of the local magnetization, a step structure contributed by spin-up atoms alone is formed with a big slope from a negative to positive value.

\section{Acknowledgements}
This work was supported by Qianjiang River Fellow Fund 2008R10029, NSF of China under Grant No. 10704066, 10974181, and Program for Innovative Research Team in Zhejiang Normal University.
We thank Lin Na for critical reading of the manuscript.
The calculations were performed using the ALPS libraries.


\begin{thebibliography}{99}
\bibitem{review paper}
        M. Lewenstein, A. Sanpera, V. Ahufinger, B. Damski, A. Sen (De), and U. Sen, Adv. Phys. {\bf 56}, 243 (2007);
        I. Bloch, J. Dalibard, and W. Zwerger, Rev. Mod. Phys. {\bf 80}, 885 (2008).
\bibitem{Kohl}
        M. K\"ohl, H. Moritz, T. St\"oferle, K. G\"unter, and T. Esslinger, Phys. Rev. Lett. {\bf 94}, 080403 (2005).
\bibitem{Chin}
        J. K. Chin, D. E. Miller, Y. Liu, C. Stan, W. Setiawan, C. Sanner, K. Xu, and W. Ketterle, Nature (London) {\bf 443}, 961 (2006).
\bibitem{Regal}
        C. A. Regal, M. Greiner, and D. S. Jin, Phys. Rev. Lett. {\bf 92}, 040403 (2004);
        G. B. Partridge, K. E. Strecker, R. I. Kamar, M. W. Jack, and R. G. Hulet, Phys. Rev. Lett. {\bf 95}, 020404 (2005).
\bibitem{Zwierlein}
        M. W. Zwierlein, A. Schirotzek, C. H. Schunck, and W. Ketterle, Science {\bf 311}, 492 (2006);
        G. B. Partridge, W. Li, R. I. Kamar, Y.A. Liao, and R. G. Hulet, Science {\bf 311}, 503 (2006);
        M. W. Zwierlein, C. H. Schunck, A. Schirotzek, and W. Ketterle, Nature (London) {\bf 442}, 54 (2006).
\bibitem{Machida1}
        M. Machida, S. Yamada, Y. Ohashi, and H. Matsumoto, Phys. Rev. A {\bf 74}, 053621 (2006);
        M. Machida, S. Yamada, M. Okumura, Y. Ohashi, and H. Matsumoto, Phys. Rev. A {\bf 77}, 053614 (2008).
\bibitem{Molina}
        R. A. Molina, J. Dukelsky, and P. Schmitteckert, Phys. Rev. Lett. {\bf 99}, 080404 (2007).
\bibitem{gaoprl98}
        G. Xianlong, M. Rizzi, Marco Polini, Rosario Fazio, M. P. Tosi, V. L. Campo, Jr., and K.
        Capelle, Phys. Rev. Lett {\bf 98}, 030404 (2007).
\bibitem{Rigol}
        M. Rigol, A. Muramatsu, G. G. Batrouni, and R. T. Scalettar, Phys. Rev. Lett. {\bf 91}, 130403 (2003).
\bibitem{Rigol2004}
        M. Rigol and A. Muramatsu, Phys. Rev. A {\bf 69}, 053612 (2004);
        Opt. Commun. {\bf 243}, 33 (2004).
\bibitem{Astrakhardik}
        G. E. Astrakharchik, D. Blume, S. Giorgini, and L. P. Pitaevskii, Phys. Rev. Lett. {\bf 93}, 050402 (2004).
\bibitem{Casula}
        M. Casula, D. M. Ceperley, and E. J. Mueller, Phys. Rev. A {\bf 78}, 033607 (2008).
\bibitem{exact diagnolization}
        T. Husslein, W. Fettes, and I. Morgenstern, Int. J. Mod. Phys. C {\bf 8}, 397 (1997).
\bibitem{Nikkarila}
        J.-P. Nikkarila, M. Koskinen, S. M. Reimann, and M. Manninen, New J. Phys. {\bf 10}, 063013 (2008);
        J.-P. Nikkarila, M. Koskinen, and M. Manninen, Eur. Phys. J. B {\bf 64}, 95 (2008).
\bibitem{mean-field theory}
        T. N. De Silva and E. J. Mueller, \pra {\bf 73}, 051602(R) (2006);
        M. Haque and H. T. C. Stoof, \pra {\bf 74}, 011602(R) (2006).
\bibitem{Recati06}
        A. Recati, I. Carusotto, C. Lobo, and S. Stringari, Phys. Rev. Lett {\bf 97}, 190403 (2006).
\bibitem{Hao}
        Y. Hao, Y. Zhang, J. Q. Liang, and S. Chen, Eur. Phys. J. D {\bf 44}, 541 (2007).
\bibitem{Roth}
        L. Salasnich, B. Pozzi, A. Parola, and L. Reatto, J. Phys. B: At. Mol. Opt. Phys. {\bf 33}, 3943 (2000);
        A. Amoruso, I. Meccoli, A. Minguzzi, and M. P. Tosi, Eur. Phys. J. D {\bf 8}, 361 (2000);
        R. Roth and H. Feldmeier, J. Phys. B: At. Mol. Opt. Phys. {\bf 34}, 4629 (2001).
\bibitem{Snoek}
        M. Snoek, I. Titvinidze, C. T\"{o}ke, K. Byczuk, and W. Hofstetter, New. J. Phys. {\bf 10}, 093008 (2008).
\bibitem{An}
        J. An and C.D. Gong, Phys. Rev. A {\bf 79}, 063607 (2009).
\bibitem{Karpiuk}
        T. Karpiuk, M. Brewczyk, and K. Rzazewski, Phys. Rev. A {\bf 69}, 043603 (2004).
\bibitem{Tezuka}
        M. Tezuka and M. Ueda, Phys. Rev. Lett. {\bf 100}, 110403 (2008).
\bibitem{Abedinpour}
        S. H. Abedinpour, M. R. Bakhtiari, G. Xianlong, M. Polini, M. Rizzi, and M. P. Tosi, Laser Phys. {\bf 17}, 162 (2007).
\bibitem{jaksch_98}
        D. Jaksch, C. Bruder, J. I. Cirac, C. W. Gardiner, and P. Zoller, \prl {\bf 81}, 3108 (1998);
        W. Hofstetter, J. I. Cirac, P. Zoller, E. Demler, and M. D. Lukin, \prl {\bf 89}, 220407 (2002).
\bibitem{moritz_2005}
        H. Moritz, T. St\"oferle, K. G\"unter, M. K\"ohl, and T. Esslinger, Phys. Rev. Lett. {\bf 94}, 210401 (2005).
\bibitem{equalmass}
        G. D. Lin, W. Yi, and L. M. Duan, Phys. Rev. A {\bf 74}, 031604(R) (2006);
        M. Iskin and C. A. R. S\'{a} de Melo, Phys. Rev. Lett. {\bf 97}, 100404 (2006);
        M. M. Parish, F. M. Marchetti, A. Lamacraft, and B. D. Simons, Phys. Rev. Lett. {\bf 98}, 160402 (2007);
        C. H. Pao, S. T. Wu, and S. K. Yip, Phys. Rev. A {\bf 76}, 053621 (2007).
\bibitem{Iskin}
        M. Iskin and C. J. Williams, Phys. Rev. A {\bf 77}, 013605 (2008); S. A. Silotri, e-print arXiv:0909.1561.
\bibitem{Blume}
        D. Blume, Phys. Rev. A {\bf 78}, 013613 (2008)
\bibitem{Jin}
        B. DeMarco and D. S. Jin, Science {\bf 285}, 1703 (1999).
\bibitem{Jordens}
        R. J\"{o}rdens, N. Strohmaier, K. G\"{u}nter, H. Moritz, and T. Esslinger, Nature (London) {\bf 455}, 204 (2008);
        U. Schneider, L. Hackerm\"{u}ller, S. Will, Th. Best, I. Bloch, T. A. Costi, R. W. Helmes, D. Rasch, and A. Rosch,
        Science {\bf 322}, 1520 (2008).
\bibitem{LiebWu}
        E. H. Lieb and F. Y. Wu, Phys. Rev. Lett {\bf 20}, 1445 (1968).
\bibitem{KocharianPRB}
        A. N. Kocharian, C. Yang, and Y. L. Chiang, Phys. Rev. B {\bf 59}, 7458 (1999).
\bibitem{gaoprb78}
        G. Xianlong, \prb {\bf 78}, 085108 (2008).
\bibitem{Liu}
        X.J. Liu, P. D. Drummond, and H. Hu, Phys. Rev. Lett. {\bf 94}, 136406 (2005).
\bibitem{Heiselberg}
        H. Heiselberg, Phys. Rev. A {\bf 74}, 033608 (2006).
\bibitem{Campo}
        V. L. Campo and K. Capelle, Phys. Rev. A {\bf 72}, 061602(R) (2005).
\bibitem{Albuquerque}
        F. Alet, S. Wessel, and M. Troyer, Phys. Rev. E {\bf 71}, 036706 (2005);
        F. Albuquerque et al., J. Magn. Magn. Mater. {\bf 310}, 1187 (2007);
        http://alps.comp-phys.org.
\bibitem{Soffing}
       S. A. S\"{o}ffing, M. Bortz, I. Schneider, A. Struck, M. Fleischhauer, and S. Eggert, Phys. Rev. B {\bf 79}, 195114 (2009).
\bibitem{MachidaPRB}
        M. Machida, M. Okumura, S. Yamada, T. Deguchi, Y. Ohashi, and H. Matsumoto, Phys. Rev. B {\bf 78}, 235117 (2008).
\end{thebibliography}
\end{document}